\documentclass[traditabstract]{aa}
\usepackage{graphics}
\usepackage{graphicx}
\usepackage{txfonts}
\usepackage{afterpage}
\usepackage{longtable,lscape}

\def\bc{\begin{center}}
\def\ec{\end{center}}
\def\be{\begin{equation}}
\def\ee{\end{equation}}
\def\bea{\begin{eqnarray}}
\def\eea{\end{eqnarray}}

\begin{document}
\title{Interstellar polarization and grain alignment: the role of iron and silicon}

\author{ N.V.~Voshchinnikov\inst{1,2,3},
         Th.~Henning\inst{1},
         M.S.~Prokopjeva\inst{2},
         and
         H.K. Das\inst{4}
         }
\authorrunning{Voshchinnikov et al.}
\titlerunning{Interstellar polarization and grain alignment}

\institute{
Max-Planck-Institut f\"ur Astronomie, K\"onigstuhl 17, D-69117 Heidelberg, Germany,
e-mail: {\tt voshchinnikov@mpia.de}
\and
Sobolev Astronomical Institute,
St.~Petersburg University, Universitetskii prosp. 28,
           St.~Petersburg, 198504 Russia,
e-mail: {\tt nvv@astro.spbu.ru}
\and
 Isaac Newton Institute of Chile, St.~Petersburg Branch
\and
IUCAA, Post Bag 4, Ganeshkhind, Pune 411 007, India
}
\date{Received 25 July 2011  / accepted 9 March 2012 }

  \abstract{\rm  We compiled the polarimetric data for a
sample of lines of sight with known  abundances of Mg, Si, and Fe.
We correlated the degree of interstellar polarization $P$
and polarization efficiency (the ratio of $P$ to the colour
excess $E(B-V)$ or extinction $A_V$) with dust phase abundances.
We detect an anticorrelation between $P$ and the
dust phase abundance of iron {\rm in non silicate--containing
grains}
$\left [{\rm {Fe(rest)}/{H}} \right ]_{\rm d}$, a
correlation between $P$ and the  abundance of Si, and
no correlation between $P/E(B-V)$ or $P/A_V$ and dust phase
abundances.
{\rm These findings can be explained if mainly the silicate grains
aligned by the radiative mechanism
are responsible for the observed interstellar linear polarization.}

\keywords{ISM: abundances -- dust,  extinction, polarization}
      }
\maketitle

\section{Introduction}

{\rm The modelling of interstellar polarization
includes light scattering calculations for aligned non-spherical particles.
It is usually performed for particles of simple shapes}
(infinite cylinders or homogeneous spheroids)
and  the particles are very often assumed to be perfectly aligned
(Mathis \cite{m86}; Kim \& Martin \cite{km95}; Draine \& Fraisse \cite{df09}).
The reasons for these simplifications are a poor knowledge of alignment
mechanisms (see Lazarian \cite{laz09}, for a recent review)
and the impossibility of light scattering calculations for complex aggregate
particles of intermediate and large sizes
(Michel et al. \cite{mich96};
Farafonov \&  Il'in \cite{fi06};
Borghese et al. \cite{bds2};  Min \cite{min09}).

According to standard concepts (Kr\"ugel \cite{kru03}; Whittet \cite{w03}),
the alignment of interstellar grains may be magnetic or radiative.
Magnetic alignment occurs if pure iron or iron components in dust
grains interact with the magnetic field.
{\rm Radiative torque alignment (RAT alignment)
arises from an azimuthal asymmetry of
the light scattering by non-spherical particles.
Magnetic inclusions can enhance RAT alignment (Lazarian \& Hoang \cite{lh08}).
The amount of iron in dust grains can be found from dust phase abundances.
Using them, we can  estimate
the grain composition and, as a consequence,
the scattering and polarizing properties of interstellar dust. }

In this {\rm paper}, we analyse the relation between the interstellar polarization
and dust phase abundances of Mg, Si, and Fe previously compiled by
Voshchinnikov \& Henning (\cite{vh10}, VH10).
We assume that all silicon and magnesium and a part of iron
are incorporated into amorphous silicates.
We correlate  the amount of the remaining iron
as well as dust phase abundances of Mg, Si, and Fe
with the polarization degree $P$ or polarization efficiency
(the ratio of $P$ to the colour excess $E(B-V)$ or extinction $A_V$)
and draw some conclusions {\rm on grains producing interstellar polarization and
favourite alignment mechanism.}

\section{Polarization and alignment mechanisms}\label{align}

{\rm The fact that interstellar grains polarize starlight
has important implications.}
It requires the
simultaneous fulfilment of the following conditions in a given direction.
\begin{enumerate}
\item Dust grains must be non-spherical.
\item Dust grains must have sizes close to the wavelength of incident
      radiation  because big particles do not polarize the transmitted
      radiation even if they are
      perfectly orientated (Voshchinnikov et al. \cite{v00}).
\item Dust grains must have specific magnetic properties to
      interact with the interstellar magnetic field.
\item Dust grains must be aligned.
\item The direction of alignment must not coincide with the line of sight.
\item The distribution of aligned grains along the line of sight must be
quite regular to exclude the cancellation of polarization.
\end{enumerate}

{\rm The most discussed item is the alignment mechanism.}
A very popular alignment mechanism is the magnetic alignment
(Davis-Greenstein (DG) type orientation, Davis \& Greenstein \cite{dg51})
based on the paramagnetic relaxation of grain material containing about
one percent of iron impurities.
The DG mechanism requires a stronger  magnetic field than average galactic
magnetic field ($\sim 3 - 5\,\mu$G;
Heiles \& Crutcher \cite{hc05}).
Because of this problem, it has been suggested that the polarizing
grains contain small clusters of iron, iron sulfides, or iron oxides
with superparamagnetic or ferromagnetic properties
(Jones \& Spitzer \cite{js67}). This leads to an enhancement of the
imaginary part of the
magnetic susceptibility of grain material $\chi''$ {\rm by a factor 10 -- 100}
and alignment can occur through the DG mechanism.
This scenario is supported by laboratory experiments
(Djouadi et al. \cite{dj07}; Belley et al. \cite{bel09}).
A significant enhancement of $\chi''$ is also possible in mixed
MgO/FeO/SiO grains (Duley \cite{dul78}) or in H$_2$O ice mantle grains
containing magnetite (Fe$_3$O$_4$) precipitates (Sorrell \cite{sor94,sor95}).

A very important factor of any alignment mechanism is grain rotation.
The faster it is, the more effective the grain alignment should be.
The DG mechanism assumes thermally rotating grains. Purcell~(\cite{pur79})
suggested a mechanism of supra-thermal spin alignment
(``pinwheel'' mechanism)
where the grains are spun up to very high velocities
as a result of the desorption of H$_2$ molecules from their surfaces.

Fast rotation can also arise because of radiation torques when
asymmetrical grains scatter the radiation (Dolginov et al. \cite{dgs79};
Draine \& Weingartner \cite{dw97}), which can lead to  grain alignment
in an anisotropic radiation field.
Radiation torque {\rm is a manifestation} of the
radiation pressure force possessing a transversal component directed
perpendicular to the direction of the incident radiation.
This component was measured in laboratory experiments
(Abbas et al. \cite{abbas04}; Krau\ss\, \& Wurm \cite{kw04}).
It can be easily calculated for cylindrical (Voshchinnikov \& Il'in \cite{vi83})
and spheroidal (Voshchinnikov \cite{v90}; Il'in \& Voshchinnikov \cite{iv98})
grains and estimated for fluffy aggregates (Kimura \& Mann \cite{km98}).
The transversal component is larger for dielectric particles and plays an
important role in the motion of interplanetary and circumstellar grains
(Il'in \& Voshchinnikov \cite{iv98}; Kimura et al. \cite{kom02};
Kocifaj \& Kla\u{c}ka \cite{kk04}).

{\rm The theory of RAT alignment is well developed (Lazarian \& Hoang \cite{lh07}).
Recent observations of interstellar polarization in the vicinity of luminous stars
(Andersson \& Potter \cite{ap10}; Matsumura et al. \cite{mat11};
Andersson et al. \cite{and11}) have been used for confirmation of the RAT
alignment mechanism.
However, the discussed models are phenomenological, they are not based on
correct light scattering calculations of interstellar polarization.
One of the reasons is that the alignment function for the RAT mechanism is unknown.
{\rm Another reason is a requirement of advanced light scattering methods
because fast rotation can only occur for grains of very specific (helical) shape
(Dolginov et al. \cite{dgs79}; Lazarian \& Hoang \cite{lh07}).}
This is highly improbable from the point of view of grain growth
in the interstellar medium.

In any event, since both magnetic alignment and radiative alignment
depend on iron inclusions,
we can expect that polarization and/or polarization efficiency should
increase with the growth of iron fraction in dust grains.
A goal of our investigation is to check this suggestion using
the available data on interstellar polarization and element
abundances.}

\section{Results and discussion}\label{res}

{\rm We compiled the polarimetric data for a
sample of 196 targets with known dust phase abundances collected in VH10.
Polarimetric observations of 13 stars in the V~band
were performed by one of the authors (HKD) in October 2011
at a 2m telescope of the  IUCAA Girawali Observatory (Pune, India).
The polarization degree $P$ in percent with 1$\sigma$ error
and the corresponding reference are
given in Table~\ref{tpp}\footnote{available online at the CDS database}
(columns 9 and 13)
in the appendix. The total number of stars with measured polarization is 172.
Columns 10 and 11 contain
the ratios of $P$ to the colour excess $E(B-V)$ and visual extinction
$A_V$\footnote{\rm Ratios of total to selective extinction  $R_V$
were taken from papers of
Fitzpatrick \& Massa~(\cite{fm07}), Valencic et al.~(\cite{vcg04}),
Wegner~(\cite{ww02}, \cite{ww03}), and Patriarchi et al.~(\cite{pmp03}).},
which characterizes the polarization
\begin{figure}[htb]
\bc
\resizebox{\hsize}{!}{\includegraphics{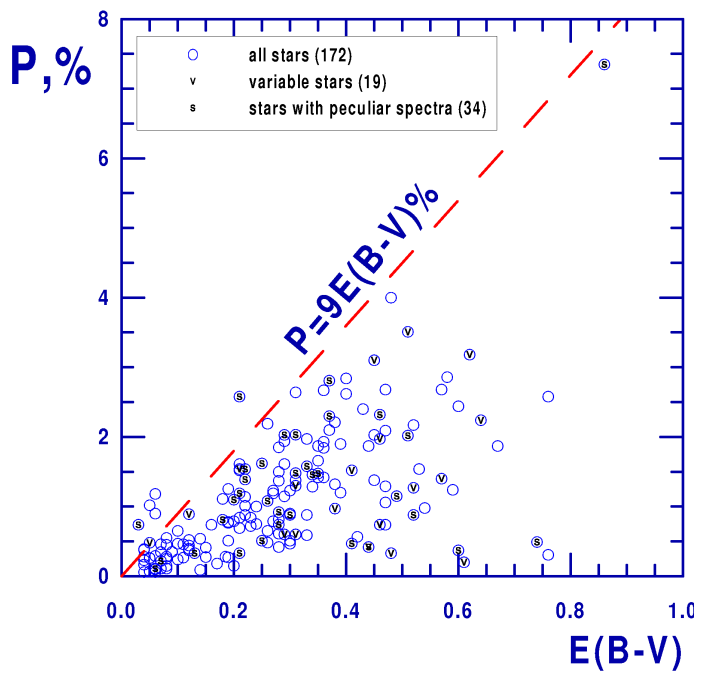}}
\caption{Polarization according to colour excess for 172 stars from
Table~\ref{tpp}. Variable stars and stars with peculiar spectra are marked.
The straight line shows the empirical upper limit for interstellar polarization
(Serkowski et al., \cite{smf75}).}
\label{p-ebv}\ec
\end{figure}
efficiency of the interstellar medium in a given direction.
In Table~\ref{tpp}, the extinction $A_V$ and the ratio $P/E(B-V)$
were calculated assuming 1$\sigma$ error for $E(B-V)$ equal to 0.01.
}

{\rm  We marked stars with a peculiar spectrum (`s') and variable stars (`v').
For these, a part of the observed polarization may be intrinsic.
However, the positions of these stars on the diagram $P$ vs. $E(B-V)$
do not look unusual (see Fig.~\ref{p-ebv}).
{\rm In the direction of variable stars or stars with peculiar spectrum
where the interstellar polarization was determined from polarization
of surrounding stars,  we used this interstellar polarization (labelled `i').}
The polarimetric data are mainly related to the visual part of spectrum
(V band). For stars marked `m' the value of the maximum polarization
$P_{\max}$ is given. The major part of observations was taken from the
catalogues of Heiles (\cite{hei00}), Serkowski et al. (\cite{smf75}), and
Efimov (\cite{ef09}). In the latter case, we used $P_{\max}$ obtained
for the Serkowski curve approximated according to
Whittet et al. (\cite{wetal92}).

Voshchinnikov \& Henning (\cite{vh10}) found a sharp
distinction in abundances of Mg, Si, and Fe
for sightlines located at low ($|b|<30\degr$) and high ($|b|>30\degr$)
galactic latitudes.
For high-latitude stars the ratios Mg/Si and Fe/Si in dust are close to 1.5.
For disk stars these ratios are reduced to $\sim 1.2$ and $\sim 1.05$
for Mg and Fe, respectively. The derived numbers indicate that
the dust grains  cannot be just a mixture of only
olivine (Mg$_{\rm 2x}$Fe$_{\rm 2-2x}$SiO$_4$)
and pyroxene (Mg$_{\rm y}$Fe$_{\rm 1-y}$SiO$_3$)
silicates (here $0 \leq x, y \leq 1$).
Some amount of magnesium or iron (or both) should be embedded into
other materials.

Based on the discussion of alignment mechanisms (Sect.~\ref{align}),
we suggest that the interstellar polarization is probably be related to
the amount of iron in dust grains.
We assume that {\it all silicon and all magnesium
are embedded into amorphous silicates of olivine composition}
(Mg$_{2x}$Fe$_{2-2x}$SiO$_4$,
where $x = [{\rm {Mg}/{H}}]_{\rm d}/(2 [{\rm {Si}/{H}}]_{\rm d})$
as is a part of iron.
The remaining part of Fe can be found as
\be
\left [{\rm {Fe(rest)}/{H}} \right ]_{\rm d}  =
\left [{\rm {Fe}/{H}} \right ]_{\rm d}  -
(2\,\left [{\rm {Si}/{H}} \right ]_{\rm d} -
\left [{\rm {Mg}/{H}} \right ]_{\rm d}).   \label{fer}
\ee
\begin{figure}[htb]
\bc
\resizebox{\hsize}{!}{\includegraphics{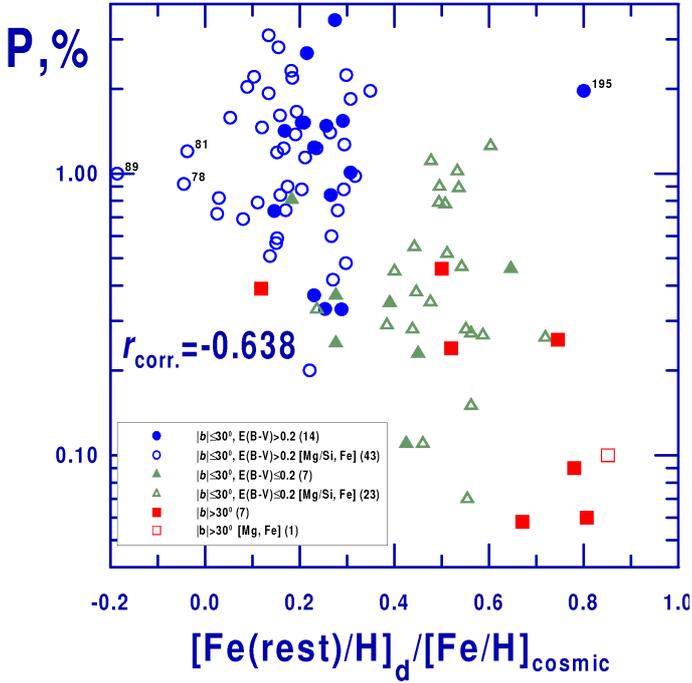}}
\caption{Polarization of 95 stars  according to the remaining dust phase abundance
of Fe as given by Eq.~(\ref{fer}) normalized to the cosmic abundance of iron.
Halo stars with $|b| > 30\degr$ and disk stars with $|b|\leq 30\degr$
and low ($E(B-V) \leq 0.2$) and high ($E(B-V) > 0.2$) reddening
are shown with different symbols.
Filled symbols correspond to sightlines where the abundances of three
elements (Mg, Si, and Fe) are measured.
{\rm Open symbols correspond to sightlines where the abundances of
two elements (Fe and Mg or Si) are known.}
The number of stars is indicated in parentheses in the legend.}
\label{f-pfe}\ec
\end{figure}
It is expected to be in the form of various iron oxides
(FeO, Fe$_2$O$_3$, Fe$_3$O$_4$) and/or metallic iron, which confers
magnetic properties to the grains.
}
{\rm

\begin{figure}[htb]
\bc
\resizebox{\hsize}{!}{\includegraphics{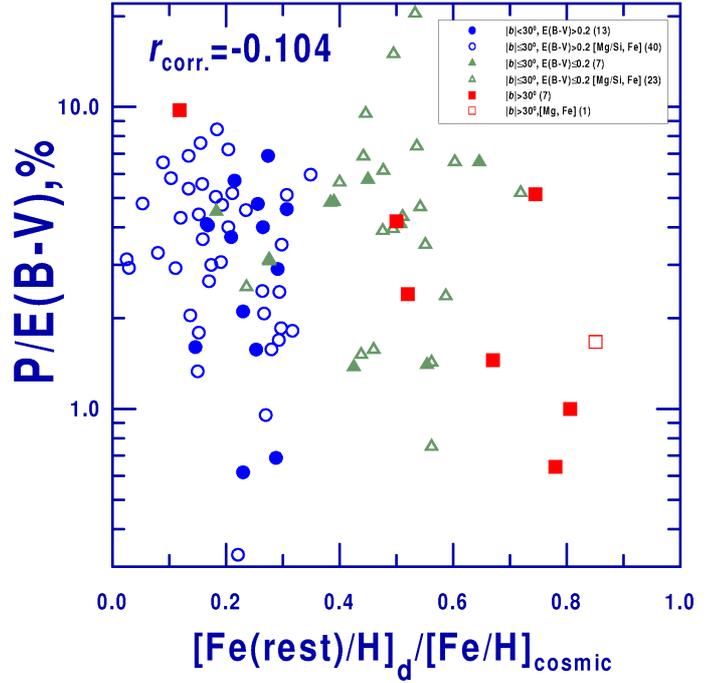}}
\caption{Same as in Fig.~\ref{f-pfe}, but now for
polarization efficiency $P/E(B-V)$.}
\label{f-pebvfe}\ec
\end{figure}

Our choice of Mg-rich silicates with the olivine stoichiometry
is based on the theoretical and observational studies of dust composition
in circumstellar environments (Gail \cite{g10}; Molster et al. \cite{mol10};
Sturm et al. \cite{sturm10}) and on the interpretation of the observed
interstellar silicate absorption profiles (Min et al. \cite{min07};
van Breemen et al. \cite{vanb11}).
Probably the single exception is the oxygen-rich AGB star T Cep,
whose peak wavelength in the IR features suggests the presence of
Fe-rich silicates (Niyogi et al. \cite{nst11}).

\setcounter{table}{1}
\begin{table*}[htb]
\begin{center}
\caption{Pearson correlation coefficients between polarization, extinction, and
element abundances in the dust phase.}
\begin{tabular}{lrrl} \hline\hline
\noalign{\smallskip}
~~~~~~~~~~~~Parameters and Figure & $N_{\rm stars}$ & $r_{\rm corr.}~$ & ~~~~~~~~~~~~Comment \\
\noalign{\smallskip}
\hline
\noalign{\smallskip}
$\log P$        vs. $\left [{\rm {Fe(rest)}/{H}} \right ]_{\rm d}/[{\rm Fe/H}]_{\rm cosmic}$, Fig.~\ref{f-pfe}     &   91  & --0.638 & Polarizing grains are not iron-rich non-silicates \\
$\log P/E(B-V)$ vs. ${\left [{\rm {Fe(rest)}/{H}} \right ]_{\rm d}}/{[{\rm Fe/H}]_{\rm cosmic}}$, Fig.~\ref{f-pebvfe}  &   91  & --0.104 & No correlation \\
$\log P/A_V$    vs. $\left [{\rm {Fe(rest)}/{H}} \right ]_{\rm d}/[{\rm Fe/H}]_{\rm cosmic}$                       &   69  & ~0.137      & "              \\
\noalign{\smallskip}
$\log P$        vs. $\left [{\rm {Fe}/{H}} \right ]_{\rm d}/[{\rm Fe/H}]_{\rm cosmic}$, Fig.~\ref{f-fe}            & 121   & ~0.351 & Weak correlation   \\
$\log P/E(B-V)$ vs. $\left [{\rm {Fe}/{H}} \right ]_{\rm d}/[{\rm Fe/H}]_{\rm cosmic}$                             & 121   & ~0.013 & No correlation    \\
$\log P/A_V$    vs. $\left [{\rm {Fe}/{H}} \right ]_{\rm d}/[{\rm Fe/H}]_{\rm cosmic}$                             & ~98   & --0.117 & "   \\
\noalign{\smallskip}
$\log P$        vs. $\left [{\rm {Mg}/{H}} \right ]_{\rm d}/[{\rm Mg/H}]_{\rm cosmic}$, Fig.~\ref{f-mg}            & 130   & ~0.296 & Weak correlation   \\
$\log P/E(B-V)$ vs. $\left [{\rm {Mg}/{H}} \right ]_{\rm d}/[{\rm Mg/H}]_{\rm cosmic}$                             & 130   & --0.050 & No correlation   \\
$\log P/A_V$    vs. $\left [{\rm {Mg}/{H}} \right ]_{\rm d}/[{\rm Mg/H}]_{\rm cosmic}$                             &  94   & --0.174 & "   \\
\noalign{\smallskip}
$\log P$        vs. $\left [{\rm {Si}/{H}} \right ]_{\rm d}/[{\rm Si/H}]_{\rm cosmic}$, Fig.~\ref{f-si}            & 34    & 0.736 & Polarizing grains are silicates    \\
$\log P/E(B-V)$ vs. $\left [{\rm {Si}/{H}} \right ]_{\rm d}/[{\rm Si/H}]_{\rm cosmic}$                             & 34    & 0.116 & No correlation     \\
$\log P/A_V$    vs. $\left [{\rm {Si}/{H}} \right ]_{\rm d}/[{\rm Si/H}]_{\rm cosmic}$                             & 16    & 0.013 & "     \\
\noalign{\smallskip}
\hline
\label{tab1}
\end{tabular}
\end{center}
\end{table*}

{\rm Using Eq.~(\ref{fer}), we calculated the dust phase abundance of Fe
incorporated into non-silicate materials
$\left [{\rm {Fe(rest)}/{H}} \right ]_{\rm d}$
(see column 13 in Table~\ref{tpp}). Then these abundances were
normalized to the solar (cosmic) abundance of iron
([Fe/H]$_{\sun}= 31.6$~ppm, Asplund et al. \cite{agss09})\footnote{The
normalization with [Fe/H]$_{\rm cosmic}$ was performed to
obtain convenient  units.}
and we correlated them with the polarization and polarization efficiency.
The results are shown in Figs.~\ref{f-pfe} and \ref{f-pebvfe} and in
Table~\ref{tab1} (rows 1 -- 3).
The abundances of the three elements Mg, Si, and Fe were
measured only for the 28 targets with known polarization.
They are plotted in the figures using filled symbols.
To increase the sample volume, we calculated
$\left [{\rm {Fe(rest)}/{H}} \right ]_{\rm d}$
for sightlines where the abundances of two elements
(Fe and Mg or Si) were known. In this case,
we use the average abundance of the third (lacking) element as given in
rows 2 -- 5 of Table~2 in VH10.
These data are shown by open symbols in Figs.~\ref{f-pfe} and \ref{f-pebvfe}.
Clearly, the amount of Fe in non-silicates is quite high,
especially for low-reddened ($E(B-V) \la 0.2$) and
halo ($|b| > 30\degr$) stars.
{\rm There is also a tendency  for a decline in polarization
(see Fig.~\ref{f-pfe}) with transition from high-reddened to low-reddened
stars, which likely reflects
a fall of the column density of polarizing grains.}
An exception is the sightline towards the star N~195 (CPD\,--59\,2603),
where the dust phase abundance of Si is very low.
There are also three targets (NN~78, 81, and 89)
where the calculated values of
$\left [{\rm {Fe(rest)}/{H}} \right ]_{\rm d}$
are negative.
We excluded these four stars from the  correlation analysis
(rows 1 -- 3 in Table~\ref{tab1}).}

As indicated in Fig.~\ref{f-pfe}, there is a negative correlation
between the polarization degree $P$ and the amount of remaining iron.
{\rm This is inconsistent with the common suggestion about
the great role of iron-rich grains in the production of polarization.
Because $P$ is proportional to the column density of polarizing grains,
we can conclude that {\it the increase of the iron content in non-silicate grains
does not enhance polarization}.}
These particles may be spherical, very large or very small in comparison with
radiation wavelength, or they may be less aligned.
It is also important that metallic iron or iron oxides
have high refractive indices, and particles consisting of these materials
scatter less radiation compared with (Mg,Fe)-silicates, which prevents
radiative torques.
{\rm Therefore, we can conservatively estimate that the
radiative alignment is the favourite
alignment mechanism of the interstellar grains.}

\begin{figure}[htb]
\bc
\resizebox{\hsize}{!}{\includegraphics{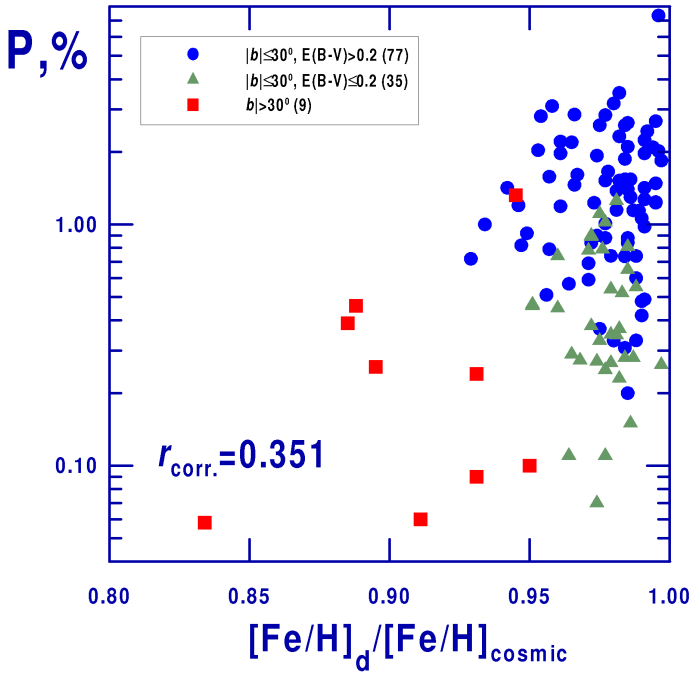}}
\caption{Interstellar polarization according to dust phase abundance of iron.
Halo stars with $|b| > 30\degr$ and disk stars with $|b|\leq 30\degr$
and low ($E(B-V) \leq 0.2$) and high ($E(B-V) > 0.2$) reddening
are shown with different symbols.
The number of stars is indicated in parentheses in the legend.}
\label{f-fe}\ec
\end{figure}

\begin{figure}[htb]
\bc
\resizebox{\hsize}{!}{\includegraphics{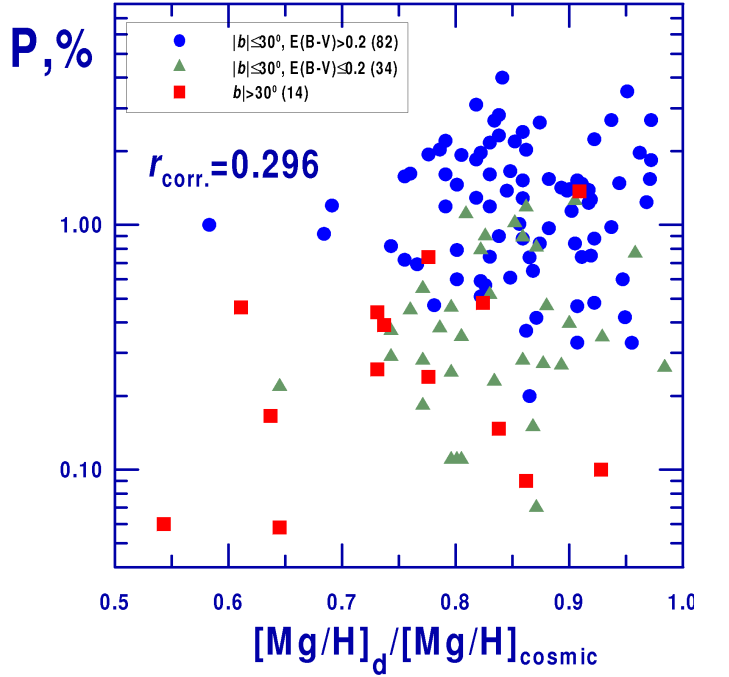}}
\caption{Same as in Fig.~\ref{f-fe}, but now for magnesium.}
\label{f-mg}\ec
\end{figure}

\begin{figure}[htb]
\bc
\resizebox{\hsize}{!}{\includegraphics{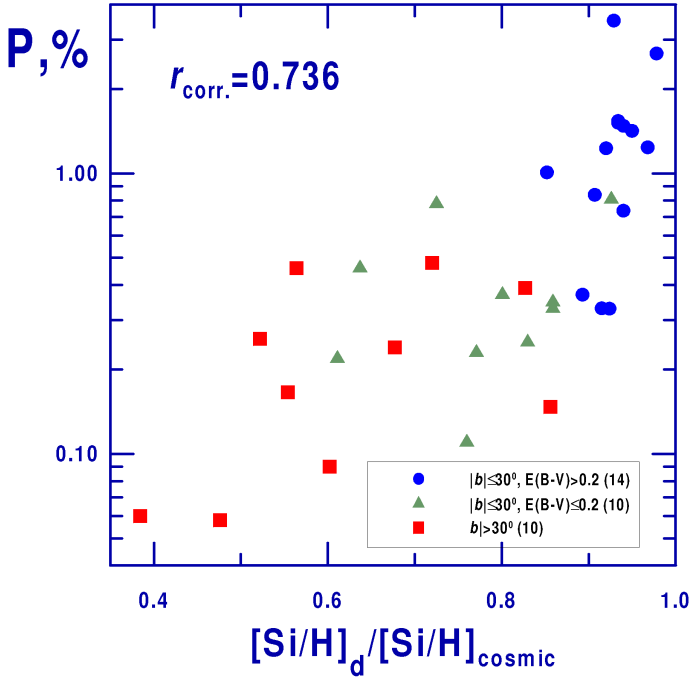}}
\caption{Same  as in Fig.~\ref{f-fe}, but now for silicon.}
\label{f-si}\ec
\end{figure}

Because in calculating $\left [{\rm {Fe(rest)}/{H}} \right ]_{\rm d}$
we removed all Si and Mg and a part of Fe from the dust phase,
we expect a positive correlation between the  polarization and the
abundances of the eliminated elements.
Figures \ref{f-fe} -- \ref{f-si} show the dependence of
interstellar polarization on dust phase abundances of iron, magnesium, and silicon.
{\rm Evidently, there is only a weak correlation between
$P$ and  $\left [{\rm {Fe}/{H}} \right ]_{\rm d}$ or
$\left [{\rm {Mg}/{H}} \right ]_{\rm d}$ and a strong correlation between
$P$ and  $\left [{\rm {Si}/{H}} \right ]_{\rm d}$.
Therefore, we can establish that  {\it polarization is more likely produced
by silicates}.}
These findings are evidence in favour of
the assumption of Mathis (\cite{m86}) that only the silicate grains are
aligned and contribute to the observed polarization, while
the carbonaceous grains are either spherical or randomly aligned.
Note that a poor alignment of carbonaceous grains compared with
silicate grains also follows from the consideration of disalignment
caused by thermal flipping or discrete charging (Weingartner \cite{wein06}).
Another verification of this suggestion is the absence of
any correlation between the polarization efficiency $P/E(B-V)$ or $P/A_V$
and dust phase abundances of elements
(see Table~\ref{tab1} and Fig.~\ref{f-pebvfe}).
This is because dust grains of all types (silicate,
carbonaceous, iron-rich, etc.) contribute to the observed extinction $A_V$ and
most likely to the colour excess $E(B-V)=A_B-A_V$, while only the silicates
seems to be responsible for the observed polarization.
Thus, the  absence of correlation between the ratio of the total to selective
extinction $R_V$ and  the wavelength of maximum polarization
$\lambda_{\max}$ observed in many cases
(e.g., Whittet et al. \cite{wetal01}; Andersson \& Potter \cite{ap07})
can be easily understood.

\section{Conclusions}

The main results of the paper can be formulated as follows.

\begin{enumerate}
\item
We compiled the polarimetric data for a
sample of 196 lines of sight with known dust phase abundances of Mg, Si, and Fe
collected in Voshchinnikov \& Henning (\cite{vh10}).
The total number of stars with measured polarization is 172.
Polarimetric observations of 13 stars  are  presented
for the first time in this paper.

\item
Assuming that all Si and Mg and a part of Fe
are incorporated into amorphous silicates of olivine composition,
we calculated the dust phase abundance of the remaining iron.
The fraction of iron not included in silicates
$\left [{\rm {Fe(rest)}/{H}} \right ]_{\rm d}$
is quite high: changing from $\sim 0.1$ of cosmic abundance of Fe
for reddened disk ($|b|<30\degr$, $E(B-V) > 0.2$) stars
to $\sim 0.8$ for halo ($|b|>30\degr$) stars.

\item
We detect an anticorrelation between $P$ and
$\left [{\rm {Fe(rest)}/{H}} \right ]_{\rm d}$ ($r_{\rm corr.}=-0.638$)
and a correlation
between $P$ and dust phase abundances of Si ($r_{\rm corr.}=0.736$).
We conclude that it is
more likely that
polarization is produced by silicates and
is not produced by iron-rich non-silicate grains.
Because silicate grains scatter more radiation than grains of
other types (carbonaceous, iron oxides, etc.), these findings
can be interpreted in favour of the radiative alignment of
interstellar grains.

\item
We found no correlation between the polarization efficiency (the ratio of $P$
to the colour excess $E(B-V)$ or extinction $A_V$) and the dust phase
abundances of Mg, Si, or Fe.
This  fact can be explained if we assume that dust grains of all types
contribute to  $A_V$ and $E(B-V)$, while only the silicates
are responsible for the observed polarization.

\end{enumerate}

{\rm A more reliable decision about the contribution of the grains
of different types to the observed linear interstellar polarization
and the grain alignment can be reached after detailed modelling.}

}
\acknowledgements{We thank
Albina Timirbaeva for assistance in collecting the observational
data, Andrei Berdyugin for sending the polarimetric data, and
Vladimir Il'in, Svitlana Zhukovska, and Alex Lazarian
for stimulating discussions.
{\rm We are grateful to the anonymous referee
for comments and suggestions.}
The work was partly supported by the grants RFBR 10-02-00593 and 11-02-92695.}


\clearpage\newpage
\setcounter{table}{0}
\begin{landscape}
\begin{table}
\section*{Appendix}
\caption[]{Polarization  of stars with known dust phase abundances.}    \label{tpp}
\bc
\begin{tabular}{llrrcrcccrccc}
\hline\hline\noalign{\smallskip}
\multicolumn{1}{l}{~~$N$} &
\multicolumn{1}{l}{~~~Star} &
\multicolumn{1}{c}{$l$} &
\multicolumn{1}{c}{~$b$} &
\multicolumn{1}{c}{Spectrum} &
\multicolumn{1}{c}{$D$, pc} &
\multicolumn{1}{c}{$E(B-V)$} &
\multicolumn{1}{c}{$A_V$} &
\multicolumn{1}{c}{$P,\,\%$} &
\multicolumn{1}{c}{$P/E(B-V),\%$} &
\multicolumn{1}{c}{$P/A_V,\%$} &
\multicolumn{1}{c}{$\left [{\rm {Fe(rest)}/{H}} \right ]_{\rm d}$} &
\multicolumn{1}{c}{Reference} \\
\multicolumn{1}{l}{~(1)} &
\multicolumn{1}{l}{~~~(2)} &
\multicolumn{1}{c}{(3)} &
\multicolumn{1}{c}{~(4)} &
\multicolumn{1}{c}{(5)} &
\multicolumn{1}{c}{(6)} &
\multicolumn{1}{c}{(7)} &
\multicolumn{1}{c}{(8)} &
\multicolumn{1}{c}{(9)} &
\multicolumn{1}{c}{(10)} &
\multicolumn{1}{c}{(11)} &
\multicolumn{1}{c}{(12)} &
\multicolumn{1}{c}{(13)} \\
\noalign{\smallskip}\hline\noalign{\smallskip}
~~1    &  HD 1383      & 119.02 & --0.89 &  B1II          &   2702   & 0.47 &   1.30   $\pm$   0.12   &  1.29    $\pm$  0.20  &   2.74    $\pm$    0.48 & 0.99   $\pm$   0.24  &         $\cdots$       &~(1)   \\
~~2v   &  HD 5394      & 123.58 & --2.15 &  B0IVpe        &    188   & 0.12 &         $\cdots$        &  0.89    $\pm$  0.07  &   7.42    $\pm$    1.20 &       $\cdots$       &   16.95   $\pm$   8.48 &~(2)   \\
~~3    &  HD 12323     & 132.91 & --5.87 &  ON9V          &   3586   & 0.21 &   0.61   $\pm$   0.08   &  1.52    $\pm$  0.18  &   7.24    $\pm$    1.20 & 2.50   $\pm$   0.64  &    6.44   $\pm$   5.15 &~(1)   \\
~~4    &  HD 13268     & 133.96 & --4.99 &  O8V           &   2391   & 0.36 &   1.09   $\pm$   0.12   &  2.67    $\pm$  0.20  &   7.42    $\pm$    0.76 & 2.46   $\pm$   0.45  &          $\cdots$      &~(1)   \\
~~5s   &  HD 13745     & 134.58 & --4.96 &  O9.7IIn       &   1900   & 0.37 &   1.14   $\pm$   0.14   &  2.81    $\pm$  0.20  &   7.59    $\pm$    0.75 & 2.46   $\pm$   0.48  &    4.89   $\pm$   6.02 &~(1)   \\
~~6    &  HD 14434     & 135.08 & --3.82 &  O6.5V         &   4108   & 0.48 &   1.23   $\pm$   0.10   &  4.00    $\pm$  0.20  &   8.33    $\pm$    0.59 & 3.24   $\pm$   0.43  &          $\cdots$      &~(1)   \\
~~7s   &  HD 15137     & 137.46 &  +7.58 &  O9.5II-IIIn   &   3300   & 0.31 &   1.10   $\pm$   0.13   &  2.032   $\pm$  0.0414&   6.55    $\pm$    0.34 & 1.85   $\pm$   0.25  &    2.81   $\pm$   7.19 &~(3)   \\
~~8    &  HD 18100     & 217.93 &--62.73 &  B5II-III      &   3100   & 0.05 &         $\cdots$        &  0.257   $\pm$  0.0367&   5.14    $\pm$    1.76 &       $\cdots$       &   23.55   $\pm$  19.70 &~(3)   \\
~~9    &  HD 21856     & 156.32 &--16.75 &  B1V           &    500   & 0.19 &         $\cdots$        &  0.78    $\pm$  0.20  &   4.11    $\pm$    1.27 &       $\cdots$       &   16.02   $\pm$  16.34 &~(1)   \\
 10    &  HD 22586     & 264.19 &--50.36 &  B2III         &   2020   & 0.06 &         $\cdots$        &  0.100   $\pm$  0.035 &   1.67    $\pm$    0.86 &       $\cdots$       &   26.88   $\pm$  11.87 &~(1)   \\
 11    &  HD 22928     & 150.28 & --5.77 &  B5III         &    160   & 0.05 &         $\cdots$        &  0.262   $\pm$  0.069 &   5.24    $\pm$    2.43 &       $\cdots$       &   22.73   $\pm$   6.33 &~(1)   \\
 12    &  HD 22951     & 158.92 &--16.70 &  B0.5V         &    320   & 0.19 &         $\cdots$        &  0.766   $\pm$  0.099 &   4.03    $\pm$    0.73 &       $\cdots$       &          $\cdots$      &~(1)   \\
 13v   &  HD 23180     & 160.36 &--17.74 &  B1IVSB        &    219   & 0.29 &   0.91   $\pm$   0.10   &  0.6     $\pm$  0.2   &   2.07    $\pm$    0.76 & 0.66   $\pm$   0.29  &    8.44   $\pm$   3.22 &~(4)   \\
 14    &  HD 23478     & 160.76 &--17.42 &  B3IV          &    240   & 0.28 &         $\cdots$        &  1.5     $\pm$  0.05  &   5.36    $\pm$    0.37 &       $\cdots$       &          $\cdots$      &~(4)   \\
 15    &  HD 24190     & 160.39 &--15.90 &  B2V           &    550   & 0.30 &         $\cdots$        &  0.51    $\pm$  0.20  &   1.70    $\pm$    0.72 &       $\cdots$       &          $\cdots$      &~(1)   \\
 16m   &  HD 24398     & 162.29 &--16.69 &  B1Ib          &    301   & 0.27 &   0.71   $\pm$   0.07   &  1.23    $\pm$  0.088 &   4.56    $\pm$    0.49 & 1.73   $\pm$   0.29  &    7.42   $\pm$   3.07 &~(5)   \\
 17im  &  HD 24534     & 163.08 &--17.14 &  O9.5pe        &    590   & 0.59 &   2.05   $\pm$   0.21   &  1.24    $\pm$  0.01  &   2.10    $\pm$    0.05 & 0.61   $\pm$   0.07  &    7.27   $\pm$   2.23 &~(6)   \\
 18    &  HD 24760     & 157.35 &--10.09 &  B0.5IV        &    165   & 0.11 &         $\cdots$        &  0.267   $\pm$  0.063 &   2.43    $\pm$    0.79 &       $\cdots$       &   18.54   $\pm$   8.26 &~(1)   \\
 19m   &  HD 24912     & 160.37 &--13.10 &  O7V           &    421   & 0.35 &   1.00   $\pm$   0.21   &  1.42    $\pm$  0.03  &   4.06    $\pm$    0.20 & 1.42   $\pm$   0.32  &    5.30   $\pm$   2.72 &~(7)   \\
 20    &  HD 27778     & 172.76 &--17.39 &  B3V           &    262   & 0.36 &   0.96   $\pm$   0.08   &  1.841   $\pm$  0.0312&   5.11    $\pm$    0.23 & 1.91   $\pm$   0.19  &    9.69   $\pm$   1.78 &~(3)   \\
 21v   &  HD 28497     & 208.78 &--37.40 &  B2Vne         &    483   & 0.05 &         $\cdots$        &  0.48    $\pm$  0.05  &   9.60    $\pm$    2.92 &       $\cdots$       &          $\cdots$      &~(2)   \\
 22m   &  HD 30614     & 144.07 &  14.04 &  O9.5Ia        &    963   & 0.29 &   0.87   $\pm$   0.13   &  1.61    $\pm$  0.02  &   5.55    $\pm$    0.26 & 1.84   $\pm$   0.29  &    4.99   $\pm$   6.67 &~(8)   \\
 23    &  HD 34029     & 162.59 &  +4.57 &  G8III+G0III   &     13   & 0.01 &         $\cdots$        &        $\cdots$       &         $\cdots$        &       $\cdots$       &   19.46   $\pm$   6.34 &       \\
 24    &  HD 34816     & 214.83 &--26.24 &  B0.5IV        &    260   & 0.05 &         $\cdots$        &        $\cdots$       &         $\cdots$        &       $\cdots$       &          $\cdots$      &       \\
 25    &  HD 34989     & 194.62 &--15.61 &  B1V           &    490   & 0.10 &   0.27   $\pm$   0.06   &  0.65    $\pm$  0.20  &   6.50    $\pm$    2.65 & 2.37   $\pm$   1.27  &          $\cdots$      &~(1)   \\
 26    &  HD 35149     & 199.16 &--17.86 &  B1V           &    295   & 0.12 &   0.39   $\pm$   0.07   &  0.37    $\pm$  0.20  &   3.08    $\pm$    1.92 & 0.94   $\pm$   0.67  &    8.73   $\pm$   7.07 &~(1)   \\
 27    &  HD 35715     & 200.09 &--17.22 &  B2IV          &    370   & 0.04 &         $\cdots$        &  0.28    $\pm$  0.05  &   7.00    $\pm$    3.00 &       $\cdots$       &          $\cdots$      &~(1)   \\
 28    &  HD 36486     & 203.86 &--17.74 &  O9.5II        &    281   & 0.08 &         $\cdots$        &  0.11    $\pm$  0.10  &   1.38    $\pm$    1.42 &       $\cdots$       &   13.44   $\pm$   4.03 &~(1)   \\
 29    &  HD 36822     & 195.40 &--12.29 &  B0.5IV-V      &    330   & 0.08 &   0.21   $\pm$   0.07   &  0.28    $\pm$  0.10  &   3.50    $\pm$    1.69 & 1.33   $\pm$   0.90  &   17.42   $\pm$   9.05 &~(1)   \\
 30m   &  HD 36861     & 195.05 &--12.00 &  O8IIIf        &    550   & 0.09 &   0.22   $\pm$   0.08   &  0.35    $\pm$  0.113 &   3.89    $\pm$    1.69 & 1.58   $\pm$   1.07  &   15.05   $\pm$   9.45 &~(9)   \\
 31v   &  HD 37021     & 209.01 &--19.38 &  B0V           &    678   & 0.48 &   2.80   $\pm$   0.18   &  0.33    $\pm$  0.08  &   0.69    $\pm$    0.18 & 0.12   $\pm$   0.04  &    9.10   $\pm$   4.05 &(10)   \\
 32    &  HD 37043     & 209.52 &--19.58 &  O9III         &    406   & 0.07 &         $\cdots$        &  0.11    $\pm$  0.02  &   1.57    $\pm$    0.51 &       $\cdots$       &   14.55   $\pm$   8.32 &~(1)   \\
 33m   &  HD 37061     & 208.92 &--19.27 &  B0.5 V        &    476   & 0.53 &   2.41   $\pm$   0.11   &  1.54    $\pm$  0.077 &   2.91    $\pm$    0.20 & 0.64   $\pm$   0.06  &    9.20   $\pm$   2.11 &~(5)   \\
 34    &  HD 37128     & 205.21 &--17.24 &  B0Iab         &    412   & 0.04 &         $\cdots$        &  0.23    $\pm$  0.02  &   5.75    $\pm$    1.94 &       $\cdots$       &   14.23   $\pm$   5.79 &~(1)   \\
 35v   &  HD 37367     & 179.04 & --1.03 &  B2 V SB       &    273   & 0.38 &   1.27   $\pm$   0.09   &  0.97    $\pm$  0.20  &   2.55    $\pm$    0.59 & 0.76   $\pm$   0.21  &          $\cdots$      &~(1)   \\
 36    &  HD 37468     & 206.82 &--17.34 &  O9.5V         &    370   & 0.06 &         $\cdots$        &  0.29    $\pm$  0.02  &   4.83    $\pm$    1.14 &       $\cdots$       &   12.13   $\pm$   9.98 &~(1)   \\
 37m   &  HD 37903     & 206.85 &--16.54 &  B1.5V         &    719   & 0.33 &   1.30   $\pm$   0.11   &  1.97    $\pm$  0.01  &   5.97    $\pm$    0.21 & 1.51   $\pm$   0.13  &   11.02   $\pm$   3.65 &~(7)   \\
 38m   &  HD 38087     & 207.07 &--16.26 &  B5V           &    315   & 0.31 &   1.80   $\pm$   0.14   &  2.64    $\pm$  0.01  &   8.52    $\pm$    0.31 & 1.47   $\pm$   0.12  &          $\cdots$      &~(8)   \\
 39    &  HD 38666     & 237.29 &--27.10 &  O9.5V         &    397   & 0.06 &         $\cdots$        &  0.060   $\pm$  0.035 &   1.00    $\pm$    0.75 &       $\cdots$       &   25.48   $\pm$   5.68 &~(1)   \\
 40m   &  HD 38771     & 214.51 &--18.50 &  B0Iab         &    221   & 0.12 &         $\cdots$        &  0.52    $\pm$  0.075 &   4.33    $\pm$    0.99 &       $\cdots$       &   16.15   $\pm$   7.42 &~(5)   \\
 41    &  HD 40111     & 183.97 &  +0.84 &  B0.5II        &    480   & 0.18 &   0.55   $\pm$   0.09   &  1.11    $\pm$  0.20  &   6.17    $\pm$    1.45 & 2.02   $\pm$   0.68  &   15.08   $\pm$   9.95 &~(1)   \\
 42    &  HD 40893     & 180.09 &  +4.34 &  B0IV:         &   2632   & 0.45 &   1.22   $\pm$   0.09   &  1.38    $\pm$  0.18  &   3.07    $\pm$    0.47 & 1.13   $\pm$   0.23  &    6.03   $\pm$   4.94 &~(1)   \\
 43m   &  HD 41117     & 189.69 & --0.86 &  B2Ia          &    909   & 0.40 &   1.25   $\pm$   0.15   &  2.84    $\pm$  0.274 &   7.10    $\pm$    0.86 & 2.28   $\pm$   0.49  &          $\cdots$      &~(5)   \\
 44s   &  HD 41161     & 164.97 & +12.89 &  O8Vn          &   1400   & 0.21 &   0.55   $\pm$   0.17   &  2.58    $\pm$  0.20  &  12.29    $\pm$    1.54 & 4.67   $\pm$   1.81  &          $\cdots$      &~(1)   \\
 45m   &  HD 42087     & 187.75 &  +1.77 &  B2.5Ib        &   1578   & 0.37 &   1.17   $\pm$   0.20   &  2.10    $\pm$  0.075 &   5.68    $\pm$    0.36 & 1.80   $\pm$   0.37  &          $\cdots$      &~(5)   \\
\noalign{\smallskip}\hline\noalign{\smallskip}
\end{tabular}
\ec
\end{table}
\end{landscape}
\setcounter{table}{0}
\begin{landscape}
\begin{table}
\caption[]{continued.}
\bc
\begin{tabular}{llrrcrcccrccc}
\hline\hline\noalign{\smallskip}
\multicolumn{1}{l}{~~$N$} &
\multicolumn{1}{l}{~~~Star} &
\multicolumn{1}{c}{$l$} &
\multicolumn{1}{c}{~$b$} &
\multicolumn{1}{c}{Spectrum} &
\multicolumn{1}{c}{$D$, pc} &
\multicolumn{1}{c}{$E(B-V)$} &
\multicolumn{1}{c}{$A_V$} &
\multicolumn{1}{c}{$P,\,\%$} &
\multicolumn{1}{c}{$P/E(B-V),\%$} &
\multicolumn{1}{c}{$P/A_V,\%$} &
\multicolumn{1}{c}{$\left [{\rm {Fe(rest)}/{H}} \right ]_{\rm d}$} &
\multicolumn{1}{c}{Reference} \\
\multicolumn{1}{l}{~(1)} &
\multicolumn{1}{l}{~~~(2)} &
\multicolumn{1}{c}{(3)} &
\multicolumn{1}{c}{~(4)} &
\multicolumn{1}{c}{(5)} &
\multicolumn{1}{c}{(6)} &
\multicolumn{1}{c}{(7)} &
\multicolumn{1}{c}{(8)} &
\multicolumn{1}{c}{(9)} &
\multicolumn{1}{c}{(10)} &
\multicolumn{1}{c}{(11)} &
\multicolumn{1}{c}{(12)} &
\multicolumn{1}{c}{(13)} \\
\noalign{\smallskip}\hline\noalign{\smallskip}
 46m   &  HD 43384     & 187.99 &  +3.53 &  B3Ia          &   1100   & 0.58 &   1.77   $\pm$   0.20   &  2.86    $\pm$  0.05  &   4.93    $\pm$    0.17 & 1.61   $\pm$   0.21  &          $\cdots$      &~(7)   \\
 47    &  HD 43818     & 188.49 &  +3.87 &  B0II          &   1623   & 0.52 &   1.80   $\pm$   0.13   &  2.17    $\pm$  0.20  &   4.17    $\pm$    0.46 & 1.21   $\pm$   0.20  &          $\cdots$      &~(1)   \\
 48s   &  HD 46056     & 206.34 & --2.25 &  O8V(n)        &   1670   & 0.49 &   1.40   $\pm$   0.08   &  1.15    $\pm$  0.20  &   2.35    $\pm$    0.46 & 0.82   $\pm$   0.19  &          $\cdots$      &~(1)   \\
 49    &  HD 46202     & 206.31 & --2.00 &  O9V           &   1670   & 0.47 &   1.47   $\pm$   0.08   &  1.06    $\pm$  0.20  &   2.26    $\pm$    0.47 & 0.72   $\pm$   0.18  &          $\cdots$      &~(1)   \\
 50s   &  HD 47839     & 202.94 &  +2.20 &  O7Ve          &    313   & 0.07 &         $\cdots$        &  0.219   $\pm$  0.132 &   3.13    $\pm$    2.33 &       $\cdots$       &          $\cdots$      &~(1)   \\
 51    &  HD 48915     & 227.23 & --8.89 &  A1V           &   3&$\!\!$--0.01&         $\cdots$        &        $\cdots$       &         $\cdots$        &       $\cdots$       &          $\cdots$      &       \\
 52    &  HD 52266     & 219.13 & --0.68 &  O9IV          &   1735   & 0.26 &   0.86   $\pm$   0.09   &  0.65    $\pm$  0.20  &   2.50    $\pm$    0.87 & 0.76   $\pm$   0.31  &          $\cdots$      &~(1)   \\
 53s   &  HD 53367     & 223.71 & --1.90 &  B0IV:e        &    780   & 0.74 &   1.76   $\pm$   0.76   &  0.49    $\pm$  0.01  &   0.66    $\pm$    0.02 & 0.28   $\pm$   0.13  &          $\cdots$      &(11)   \\
 54    &  HD 53975     & 225.68 & --2.32 &  O7.5V         &   1400   & 0.185&   0.60   $\pm$   0.07   &  0.28    $\pm$  0.20  &   1.51    $\pm$    1.16 & 0.47   $\pm$   0.39  &   13.84   $\pm$   8.66 &~(1)   \\
55i  &  HD 54662     & 224.17 & --0.78 &  O6.5V         &   1220   & 0.26 &   0.81   $\pm$   0.09   &  0.481   $\pm$  0.200 &   1.85    $\pm$    0.84 & 0.59   $\pm$   0.31  &    9.39   $\pm$   5.80 &(12)   \\
56   &  HD 57060     & 237.82 & --5.37 &  O7Iabfp       &   1870   & 0.14 &   0.70   $\pm$   0.09   &        $\cdots$       &         $\cdots$        &       $\cdots$       &   12.17   $\pm$  11.88 &       \\
57   &  HD 57061     & 238.18 & --5.54 &  O9II          &    980   & 0.10 &   0.36   $\pm$   0.07   &        $\cdots$       &         $\cdots$        &       $\cdots$       &    9.66   $\pm$   7.78 &       \\
58m  &  HD 62542     & 255.92 & --9.24 &  B5V           &    396   & 0.36 &   0.99   $\pm$   0.08   &  1.42    $\pm$  0.01  &   3.94    $\pm$    0.14 & 1.44   $\pm$   0.12  &          $\cdots$      &~(7)   \\
59   &  HD 63005     & 242.47 & --0.93 &  O6Vf          &   5200   & 0.28 &   0.83   $\pm$   0.55   &  0.417   $\pm$  0.052 &   1.49    $\pm$    0.24 & 0.50   $\pm$   0.39  &          $\cdots$      &~(1)   \\
60m  &  HD 64760     & 262.06 &--10.42 &  B0.5Ib        &    510   & 0.08 &   0.23   $\pm$   0.07   &  0.45    $\pm$  0.035 &   5.63    $\pm$    1.14 & 1.95   $\pm$   0.75  &   12.65   $\pm$  10.63 &~(5)   \\
61   &  HD 65575     & 266.68 &--12.32 &  B3IVp         &    140   & 0.05 &         $\cdots$        &        $\cdots$       &         $\cdots$        &       $\cdots$       &   22.98   $\pm$   6.09 &       \\
62s  &  HD 65818     & 263.48 &--10.28 &  B2II/IIIn     &    290   & 0.06 &         $\cdots$        &  0.110   $\pm$  0.035 &   1.83    $\pm$    0.89 &       $\cdots$       &          $\cdots$      &~(1)   \\
63   &  HD 66788     & 245.43 &  +2.05 &  O8V           &   4200   & 0.20 &   0.69   $\pm$   0.09   &  0.79    $\pm$  0.10  &   3.95    $\pm$    0.70 & 1.14   $\pm$   0.30  &   15.62   $\pm$   9.75 &~(1)   \\
64   &  HD 66811     & 255.98 & --4.71 &  O5Ibnf        &    330   & 0.05 &   0.25   $\pm$   0.09   &        $\cdots$       &         $\cdots$        &       $\cdots$       &   29.64   $\pm$  21.74 &       \\
65   &  HD 68273     & 262.80 & --7.68 &  WC8+O9I       &    350   & 0.03 &         $\cdots$        &        $\cdots$       &         $\cdots$        &       $\cdots$       &   17.66   $\pm$  12.07 &       \\
66   &  HD 69106     & 254.52 & --1.33 &  B0.5II        &   3076   & 0.20 &   0.61   $\pm$   0.12   &  0.15    $\pm$  0.1   &   0.75    $\pm$    0.54 & 0.25   $\pm$   0.21  &   17.76   $\pm$   7.60 &~(1)   \\
67   &  HD 71634     & 273.32 &--11.52 &  B5III         &    400   & 0.13 &         $\cdots$        &        $\cdots$       &         $\cdots$        &       $\cdots$       &          $\cdots$      &       \\
68s  &  HD 72754     & 266.83 & --5.82 &  B2Ia:pshe     &    690   & 0.36 &         $\cdots$        &        $\cdots$       &         $\cdots$        &       $\cdots$       &          $\cdots$      &       \\
69m  &  HD 73882     & 260.18 &  +0.64 &  O8.5V         &    759   & 0.67 &   2.31   $\pm$   0.09   &  1.87    $\pm$  0.01  &   2.79    $\pm$    0.06 & 0.81   $\pm$   0.04  &          $\cdots$      &~(7)   \\
70m  &  HD 74375     & 275.82 &--10.86 &  B1.5III       &    440   & 0.14 &   0.53   $\pm$   0.19   &  0.54    $\pm$  0.035 &   3.86    $\pm$    0.53 & 1.02   $\pm$   0.43  &          $\cdots$      &~(5)   \\
71   &  HD 75309     & 265.86 & --1.90 &  B2Ib/II       &   2924   & 0.28 &   0.95   $\pm$   0.09   &  0.61    $\pm$  0.10  &   2.18    $\pm$    0.43 & 0.64   $\pm$   0.17  &          $\cdots$      &~(1)   \\
72m  &  HD 79186     & 267.36 &  +2.25 &  B5Ia          &    980   & 0.40 &   1.28   $\pm$   0.26   &  2.62    $\pm$  0.179 &   6.55    $\pm$    0.61 & 2.04   $\pm$   0.55  &          $\cdots$      &~(5)   \\
73   &  HD 79351     & 277.69 & --7.37 &  B2IV-V        &    140   & 0.10 &         $\cdots$        &        $\cdots$       &         $\cdots$        &       $\cdots$       &          $\cdots$      &       \\
74s  &  HD 88115     & 285.32 & --5.53 &  B1.5IIn       &   3654   & 0.20 &         $\cdots$        &  1.10    $\pm$  0.10  &   5.50    $\pm$    0.78 &       $\cdots$       &          $\cdots$      &~(1)   \\
75   &  HD 90087     & 285.16 & --2.13 &  B2/B3III      &   2716   & 0.30 &   0.98   $\pm$   0.12   &  1.23    $\pm$  0.10  &   4.10    $\pm$    0.47 & 1.25   $\pm$   0.26  &    5.25   $\pm$   5.91 &~(1)   \\
76   &  HD 91316     & 234.89 & +52.77 &  B1Iab         &   1754   & 0.04 &         $\cdots$        &  0.166   $\pm$  0.025 &   4.15    $\pm$    1.66 &       $\cdots$       &          $\cdots$      &~(1)   \\
77   &  HD 91597     & 286.86 & --2.37 &  B7/B8IV/V     &   6400   & 0.27 &   1.33   $\pm$   0.13   &  1.19    $\pm$  0.10  &   4.41    $\pm$    0.53 & 0.89   $\pm$   0.16  &    4.81   $\pm$   5.20 &~(1)   \\
78s  &  HD 91651     & 286.55 & --1.72 &  O9V:n         &   2964   & 0.28 &   0.98   $\pm$   0.10   &  0.92    $\pm$  0.10  &   3.29    $\pm$    0.47 & 0.94   $\pm$   0.20  &  --1.41   $\pm$   6.24 &~(1)   \\
79s  &  HD 91824     & 285.70 &  +0.07 &  O7V((f))      &   2910   & 0.25 &   0.77   $\pm$   0.08   &  1.620   $\pm$  0.164 &   6.48    $\pm$    0.92 & 2.10   $\pm$   0.44  &          $\cdots$      &~(1)   \\
80   &  HD 91983     & 285.88 &  +0.05 &  B1III         &   2910   & 0.29 &   0.93   $\pm$   0.12   &  1.94    $\pm$  0.40  &   6.69    $\pm$    1.61 & 2.09   $\pm$   0.70  &          $\cdots$      &~(1)   \\
81   &  HD 92554     & 287.60 & --2.02 &  O5III         &   6795   & 0.39 &   1.15   $\pm$   0.15   &  1.20    $\pm$  0.10  &   3.08    $\pm$    0.34 & 1.04   $\pm$   0.22  &  --1.20   $\pm$   8.16 &~(1)   \\
82   &  HD 93030     & 289.60 & --4.90 &  B0V           &    140   & 0.04 &   0.10   $\pm$   0.06   &  0.38    $\pm$  0.10  &   9.50    $\pm$    4.88 & 3.91   $\pm$   3.33  &   14.08   $\pm$   9.82 &~(1)   \\
83   &  HD 93205     & 287.57 & --0.71 &  O3V           &   3187   & 0.38 &   1.24   $\pm$   0.12   &  2.21    $\pm$  0.11  &   5.82    $\pm$    0.44 & 1.79   $\pm$   0.27  &    3.25   $\pm$   4.50 &(13)   \\
84is &  HD 93222     & 287.74 & --1.02 &  O7III((f))    &   2201   & 0.33 &   1.67   $\pm$   0.12   &  1.58    $\pm$  0.15  &   4.79    $\pm$    0.60 & 0.95   $\pm$   0.16  &    1.68   $\pm$   5.35 &(13)   \\
85i  &  HD 93521     & 183.14 & +62.15 &  O9Vp          &   1760   & 0.04 &         $\cdots$        &  0.058   $\pm$  0.01  &   1.45    $\pm$    0.61 &       $\cdots$       &   21.18   $\pm$  21.01 &(14)   \\
86   &  HD 93843     & 228.24 & --0.90 &  O6III         &   2548   & 0.27 &   1.05   $\pm$   0.15   &  0.79    $\pm$  0.10  &   2.93    $\pm$    0.48 & 0.75   $\pm$   0.20  &    3.51   $\pm$   5.16 &~(1)   \\
87   &  HD 94493     & 289.01 & --1.18 &  B0.5Iab       &   3888   & 0.23 &   0.82   $\pm$   0.14   &  0.72    $\pm$  0.10  &   3.13    $\pm$    0.57 & 0.88   $\pm$   0.27  &    0.79   $\pm$   5.07 &~(1)   \\
88   &  HD 99857     & 294.78 & --4.94 &  B1Ib          &   3058   & 0.33 &   1.10   $\pm$   0.13   &  0.59    $\pm$  0.10  &   1.79    $\pm$    0.36 & 0.54   $\pm$   0.16  &    4.79   $\pm$   5.93 &~(1)   \\
89   &  HD 99890     & 291.75 &  +4.43 &  B0.5V:        &   3070   & 0.24 &   0.72   $\pm$   0.10   &  1.00    $\pm$  0.10  &   4.17    $\pm$    0.59 & 1.39   $\pm$   0.34  &  --5.89   $\pm$  10.60 &~(1)   \\
90   &  HD 100340    & 258.85 & +61.23 &  B1V           &   3000   & 0.04 &         $\cdots$        &  0.39    $\pm$  0.1   &   9.75    $\pm$    4.94 &       $\cdots$       &    3.74   $\pm$   7.57 &(15)   \\
\noalign{\smallskip}\hline\noalign{\smallskip}
\end{tabular}
\ec
\end{table}
\end{landscape}
\setcounter{table}{0}
\begin{landscape}
\begin{table}
\caption[]{continued.}
\bc
\begin{tabular}{llrrcrcccrccc}
\hline\hline\noalign{\smallskip}
\multicolumn{1}{l}{~~$N$} &
\multicolumn{1}{l}{~~~Star} &
\multicolumn{1}{c}{$l$} &
\multicolumn{1}{c}{~$b$} &
\multicolumn{1}{c}{Spectrum} &
\multicolumn{1}{c}{$D$, pc} &
\multicolumn{1}{c}{$E(B-V)$} &
\multicolumn{1}{c}{$A_V$} &
\multicolumn{1}{c}{$P,\,\%$} &
\multicolumn{1}{c}{$P/E(B-V),\%$} &
\multicolumn{1}{c}{$P/A_V,\%$} &
\multicolumn{1}{c}{$\left [{\rm {Fe(rest)}/{H}} \right ]_{\rm d}$} &
\multicolumn{1}{c}{Reference} \\
\multicolumn{1}{l}{~(1)} &
\multicolumn{1}{l}{~~~(2)} &
\multicolumn{1}{c}{(3)} &
\multicolumn{1}{c}{~(4)} &
\multicolumn{1}{c}{(5)} &
\multicolumn{1}{c}{(6)} &
\multicolumn{1}{c}{(7)} &
\multicolumn{1}{c}{(8)} &
\multicolumn{1}{c}{(9)} &
\multicolumn{1}{c}{(10)} &
\multicolumn{1}{c}{(11)} &
\multicolumn{1}{c}{(12)} &
\multicolumn{1}{c}{(13)} \\
\noalign{\smallskip}\hline\noalign{\smallskip}
~91   &  HD 103779    & 296.85 & --1.02 &  B0.5II        &   3061   & 0.21 &   0.69   $\pm$   0.12   &  0.690   $\pm$  0.005 &   3.29    $\pm$    0.18 & 1.00   $\pm$   0.19  &    2.54   $\pm$   5.59 &~(1)   \\
~92   &  HD 104705    & 297.45 & --0.34 &  B0.5III       &   2082   & 0.28 &   1.22   $\pm$   0.10   &  0.820   $\pm$  0.1   &   2.93    $\pm$    0.46 & 0.67   $\pm$   0.14  &    0.91   $\pm$   6.05 &~(1)   \\
~93   &  HD 106490    & 298.23 &  +3.79 &  B2IV          &    110   & 0.06 &         $\cdots$        &        $\cdots$       &         $\cdots$        &       $\cdots$       &   12.21   $\pm$   7.60 &       \\
~94   &  HD 108248    & 300.13 & --0.36 &  B0.5IV        &    100   & 0.20 &         $\cdots$        &        $\cdots$       &         $\cdots$        &       $\cdots$       &    3.23   $\pm$  12.66 &       \\
~95   &  HD 108639    & 300.22 &  +1.95 &  B1III         &    110   & 0.35 &         $\cdots$        &  1.87    $\pm$  0.10  &   5.34    $\pm$    0.44 &       $\cdots$       &          $\cdots$      &~(1)   \\
~96   &  HD 109399    & 301.71 & --9.88 &  B1Ib          &   1900   & 0.26 &   0.81   $\pm$   0.11   &  2.19    $\pm$  0.10  &   8.42    $\pm$    0.71 & 2.70   $\pm$   0.49  &    5.81   $\pm$   5.24 &~(1)   \\
~97ms &  HD 110432    & 301.96 & --0.20 &  B0.5IIIe      &    301   & 0.51 &   2.01   $\pm$   0.35   &  2.02    $\pm$  0.102 &   3.96    $\pm$    0.28 & 1.00   $\pm$   0.22  &          $\cdots$      &~(5)   \\
~98   &  HD 111934    & 303.20 &  +2.51 &  B2Ib          &   2525   & 0.51 &   1.25   $\pm$   0.13   &        $\cdots$       &         $\cdots$        &       $\cdots$       &          $\cdots$      &       \\
~99mv &  HD 113904    & 304.67 & --2.49 &  WC5+B0Ia      &   2660   & 0.21 &   0.76   $\pm$   0.07   &  1.54    $\pm$  0.01  &   7.33    $\pm$    0.40 & 2.02   $\pm$   0.20  &          $\cdots$      &~(7)   \\
 100ms &  HD 114886    & 305.52 & --0.83 &  O9IIIn        &   1000   & 0.29 &   0.84   $\pm$   0.08   &  2.03    $\pm$  0.13  &   7.00    $\pm$    0.69 & 2.41   $\pm$   0.37  &          $\cdots$      &~(5)   \\
 101   &  HD 115071    & 305.76 &  +0.15 &  B0.5V         &   1200   & 0.44 &         $\cdots$        &  1.87    $\pm$  0.10  &   4.25    $\pm$    0.32 &       $\cdots$       &          $\cdots$      &~(1)   \\
 102i  &  HD 116658    & 316.11 & +50.84 &  B1III-IV      &     80   & 0.14 &         $\cdots$        &  0.09    $\pm$  0.01  &   0.64    $\pm$    0.12 &       $\cdots$       &   24.66   $\pm$  21.91 &(14)   \\
 103s  &  HD 116781    & 307.05 & --0.07 &  B0IIIe        &   1492   & 0.34 &   1.40   $\pm$   0.14   &  1.46    $\pm$  0.10  &   4.29    $\pm$    0.42 & 1.04   $\pm$   0.18  &    3.79   $\pm$   5.95 &~(1)   \\
 104   &  HD 116852    & 304.88 &--16.13 &  O9III         &   4832   & 0.21 &   0.51   $\pm$   0.10   &        $\cdots$       &         $\cdots$        &       $\cdots$       &    3.38   $\pm$   6.09 &       \\
 105   &  HD 119608    & 320.35 & +43.13 &  B1Ib          &   4200   & 0.12 &         $\cdots$        &  0.440   $\pm$  0.035 &   3.67    $\pm$    0.60 &       $\cdots$       &          $\cdots$      &~(1)   \\
 106   &  HD 121263    & 314.07 & +14.19 &  B2.5IV        &    120   & 0.05 &         $\cdots$        &  0.070   $\pm$  0.035 &   1.40    $\pm$    0.98 &       $\cdots$       &   17.50   $\pm$   6.84 &~(1)   \\
 107   &  HD 121968    & 333.97 & +55.84 &  B1V           &   3800   & 0.15 &         $\cdots$        &  0.410   $\pm$  0.121 &   2.73    $\pm$    0.99 &       $\cdots$       &          $\cdots$      &~(1)   \\
 108m  &  HD 122879    & 312.26 &  +1.79 &  B0Ia          &   2265   & 0.36 &   1.13   $\pm$   0.14   &  1.93    $\pm$  0.116 &   5.36    $\pm$    0.47 & 1.70   $\pm$   0.31  &    4.23   $\pm$   6.04 &~(5)   \\
 109ms &  HD 124314    & 312.67 & --0.42 &  O6Vnf         &   1100   & 0.46 &   1.42   $\pm$   0.17   &  2.32    $\pm$  0.10  &   5.04    $\pm$    0.33 & 1.63   $\pm$   0.26  &    5.76   $\pm$   4.76 &~(5)   \\
 110   &  HD 127972    & 322.77 & +16.67 &  B1.5Vne       &     90   & 0.11 &         $\cdots$        &        $\cdots$       &         $\cdots$        &       $\cdots$       &    2.99   $\pm$   8.44 &       \\
111ms  &  HD 135591    & 320.13 & --2.64 &  O7.5IIIf      &   1250   & 0.22 &   0.79   $\pm$   0.08   &  1.54    $\pm$  0.084 &   7.00    $\pm$    0.70 & 1.96   $\pm$   0.29  &          $\cdots$      &~(5)   \\
112    &  HD 136298    & 331.32 & +13.82 &  B1.5IV        &    210   & 0.07 &         $\cdots$        &        $\cdots$       &         $\cdots$        &       $\cdots$       &    4.44   $\pm$   8.38 &       \\
113s   &  HD 137595    & 336.72 & +18.86 &  B3Vn          &    400   & 0.25 &         $\cdots$        &        $\cdots$       &         $\cdots$        &       $\cdots$       &          $\cdots$      &       \\
114    &  HD 138690    & 333.19 & +11.89 &  B2IV          &    130   & 0.07 &         $\cdots$        &        $\cdots$       &         $\cdots$        &       $\cdots$       &   19.35   $\pm$   6.92 &       \\
115ms  &  HD 141637    & 346.10 & +21.70 &  B2.5Vn        &    160   & 0.18 &   0.76   $\pm$   0.17   &  0.81    $\pm$  0.20  &   4.50    $\pm$    1.36 & 1.07   $\pm$   0.51  &    5.79   $\pm$   4.55 &~(5)   \\
116    &  HD 143018    & 347.21 & +20.23 &  B1V           &    141   & 0.07 &         $\cdots$        &  0.348   $\pm$  0.107 &   4.97    $\pm$    2.24 &       $\cdots$       &   12.33   $\pm$   4.38 &~(1)   \\
117    &  HD 143118    & 338.77 & +11.01 &  B2.5IV        &    140   & 0.02 &         $\cdots$        &        $\cdots$       &         $\cdots$        &       $\cdots$       &   18.77   $\pm$   7.01 &       \\
118s   &  HD 143275    & 350.10 & +22.49 &  B0.3IVe       &    123   & 0.21 &   0.65   $\pm$   0.09   &  0.331   $\pm$  0.036 &   1.58    $\pm$    0.25 & 0.51   $\pm$   0.13  &    7.99   $\pm$   3.30 &~(1)   \\
119m   &  HD 144217    & 353.19 & +23.60 &  B0.5V         &    163   & 0.21 &   0.57   $\pm$   0.09   &  0.84    $\pm$  0.185 &   4.00    $\pm$    1.07 & 1.47   $\pm$   0.56  &    8.37   $\pm$   1.79 &~(5)   \\
120m   &  HD 144470    & 352.76 & +22.76 &  B1V           &    183   & 0.22 &   0.82   $\pm$   0.10   &  1.14    $\pm$  0.017 &   5.18    $\pm$    0.31 & 1.40   $\pm$   0.20  &    6.67   $\pm$   3.07 &~(5)   \\
121s   &  HD 144965    & 339.04 &  +8.42 &  B2Vne         &    290   & 0.35 &         $\cdots$        &        $\cdots$       &         $\cdots$        &       $\cdots$       &          $\cdots$      &       \\
122v   &  HD 147165    & 351.33 & +17.00 &  B1IIISB,V     &    137   & 0.41 &   1.48   $\pm$   0.09   &  1.52    $\pm$  0.01  &   3.71    $\pm$    0.11 & 1.03   $\pm$   0.07  &    6.60   $\pm$   5.99 &(16)   \\
123    &  HD 147683    & 344.86 & +10.09 &  B4V           &    280   & 0.39 &         $\cdots$        &  1.90    $\pm$  0.10  &   4.87    $\pm$    0.38 &       $\cdots$       &          $\cdots$      &~(1)   \\
124mv  &  HD 147888    & 353.65 & +17.71 &  B3V:SB        &    195   & 0.51 &   2.08   $\pm$   0.11   &  3.51    $\pm$  0.01  &   6.88    $\pm$    0.15 & 1.69   $\pm$   0.09  &    8.65   $\pm$   3.88 &~(7)   \\
125m   &  HD 147933    & 353.68 & +17.70 &  B1.5V         &    118   & 0.47 &   2.07   $\pm$   0.11   &  2.68    $\pm$  0.01  &   5.70    $\pm$    0.14 & 1.29   $\pm$   0.08  &    6.78   $\pm$   1.65 &~(7)   \\
126vs  &  HD 148184    & 357.93 & +20.68 &  B1.5Ve        &    160   & 0.44 &   2.00   $\pm$   0.18   &  0.42    $\pm$  0.10  &   0.95    $\pm$    0.25 & 0.21   $\pm$   0.07  &    8.54   $\pm$   3.25 &(17)   \\
127    &  HD 148594    & 350.93 & +13.94 &  B9:V          &    134   & 0.21 &   0.65   $\pm$   0.09   &        $\cdots$       &         $\cdots$        &       $\cdots$       &          $\cdots$      &       \\
128mv  &  HD 149404    & 340.54 &  +3.01 &  O9Ia          &    908   & 0.62 &   2.19   $\pm$   0.27   &  3.18    $\pm$  0.10  &   5.13    $\pm$    0.24 & 1.45   $\pm$   0.23  &          $\cdots$      &~(5)   \\
129ms  &  HD 149757    &   6.28 & +23.59 &  O9.5Vnn       &    146   & 0.31 &   0.95   $\pm$   0.09   &  1.48    $\pm$  0.01  &   4.77    $\pm$    0.19 & 1.55   $\pm$   0.16  &    8.09   $\pm$   1.33 &~(7)   \\
130    &  HD 149881    &  31.37 & +36.23 &  B0.5III       &   2100   & 0.11 &         $\cdots$        &  0.46    $\pm$  0.03  &   4.18    $\pm$    0.65 &       $\cdots$       &   15.81   $\pm$  26.69 &~(1)   \\
131i   &  HD 151804    & 343.62 &  +1.94 &  O8Iab         &   1254   & 0.30 &   1.30   $\pm$   0.13   &  0.90    $\pm$  0.10  &   3.00    $\pm$    0.43 & 0.69   $\pm$   0.15  &    5.51   $\pm$   7.34 &(18)   \\
132    &  HD 151805    & 343.20 &  +1.59 &  B1Ib          &   6009   & 0.43 &   1.41   $\pm$   0.16   &        $\cdots$       &         $\cdots$        &       $\cdots$       &          $\cdots$      &       \\
133m   &  HD 152236    & 343.03 &  +0.87 &  B1Ia          &    612   & 0.60 &   2.24   $\pm$   0.27   &  2.44    $\pm$  0.035 &   4.07    $\pm$    0.13 & 1.09   $\pm$   0.15  &          $\cdots$      &~(5)   \\
134iv  &  HD 152590    & 344.84 &  +1.83 &  O7.5V         &   1800   & 0.46 &   1.51   $\pm$   0.17   &  0.737   $\pm$  0.100 &   1.60    $\pm$    0.25 & 0.49   $\pm$   0.12  &    4.61   $\pm$   3.71 &(12)  \\
135    &  HD 154368    & 349.97 &  +3.22 &  O9Ib          &   1046   & 0.76 &   2.53   $\pm$   0.15   &  0.308   $\pm$  0.007 &   0.41    $\pm$    0.01 & 0.12   $\pm$   0.01  &          $\cdots$      &~(1)   \\
\noalign{\smallskip}\hline\noalign{\smallskip}
\end{tabular}
\ec
\end{table}
\end{landscape}
\setcounter{table}{0}
\begin{landscape}
\begin{table}
\caption[]{continued.}
\bc
\begin{tabular}{llrrcrcccrccc}
\hline\hline\noalign{\smallskip}
\multicolumn{1}{l}{~~$N$} &
\multicolumn{1}{l}{~~~Star} &
\multicolumn{1}{c}{$l$} &
\multicolumn{1}{c}{~$b$} &
\multicolumn{1}{c}{Spectrum} &
\multicolumn{1}{c}{$D$, pc} &
\multicolumn{1}{c}{$E(B-V)$} &
\multicolumn{1}{c}{$A_V$} &
\multicolumn{1}{c}{$P,\,\%$} &
\multicolumn{1}{c}{$P/E(B-V),\%$} &
\multicolumn{1}{c}{$P/A_V,\%$} &
\multicolumn{1}{c}{$\left [{\rm {Fe(rest)}/{H}} \right ]_{\rm d}$} &
\multicolumn{1}{c}{Reference} \\
\multicolumn{1}{l}{~(1)} &
\multicolumn{1}{l}{~~~(2)} &
\multicolumn{1}{c}{(3)} &
\multicolumn{1}{c}{~(4)} &
\multicolumn{1}{c}{(5)} &
\multicolumn{1}{c}{(6)} &
\multicolumn{1}{c}{(7)} &
\multicolumn{1}{c}{(8)} &
\multicolumn{1}{c}{(9)} &
\multicolumn{1}{c}{(10)} &
\multicolumn{1}{c}{(11)} &
\multicolumn{1}{c}{(12)} &
\multicolumn{1}{c}{(13)} \\
\noalign{\smallskip}\hline\noalign{\smallskip}
136is  &  HD 155806    & 352.59 &  +2.87 &  O7.5Ve        &    860   & 0.28 &   0.69   $\pm$   0.56   &  0.742   $\pm$  0.090 &   2.65    $\pm$    0.42 & 1.07   $\pm$   1.00  &    5.38   $\pm$   6.89 &(12)  \\
137s   &  HD 156110    &  70.99 & +35.91 &  B3Vn          &    720   & 0.03 &         $\cdots$        &  0.74    $\pm$  0.20  &  24.67    $\pm$   14.89 &       $\cdots$       &          $\cdots$      &~(1)   \\
138m   &  HD 157246    & 334.64 &--11.48 &  B1Ib          &    348   & 0.06 &   0.17   $\pm$   0.05   &  0.90    $\pm$  0.1   &  15.00    $\pm$    4.17 & 5.32   $\pm$   2.04  &   15.65   $\pm$   9.15 &~(5)   \\
139    &  HD 157857    &  12.97 & +13.31 &  O7V           &   1902   & 0.43 &   1.48   $\pm$   0.13   &  2.40    $\pm$  0.20  &   5.58    $\pm$    0.59 & 1.62   $\pm$   0.28  &          $\cdots$      &~(1)   \\
140    &  HD 158926    & 351.74 & --2.21 &  B2IV          &    220   & 0.10 &         $\cdots$        &        $\cdots$       &         $\cdots$        &       $\cdots$       &   12.14   $\pm$   4.72 &       \\
141    &  HD 160578    & 351.04 & --4.72 &  B1.5III       &    142   & 0.08 &         $\cdots$        &  0.250   $\pm$  0.035 &   3.13    $\pm$    0.83 &       $\cdots$       &    8.73   $\pm$   9.70 &~(1)   \\
142ms  &  HD 164740    &   5.97 & --1.17 &  O7.5V(n)      &   1330   & 0.86 &   4.48   $\pm$   0.14   &  7.35    $\pm$  0.1   &   8.55    $\pm$    0.22 & 1.64   $\pm$   0.07  &          $\cdots$      &~(5)   \\
143m   &  HD 165024    & 343.33 &--13.82 &  B2Ib          &    250   & 0.05 &   0.17   $\pm$   0.07   &  1.02    $\pm$  0.035 &  20.40    $\pm$    4.78 & 5.96   $\pm$   2.62  &   16.83   $\pm$   8.40 &~(5)   \\
144s   &  HD 165955    & 357.41 & --7.43 &  B1Vnp         &   1640   & 0.21 &         $\cdots$        &  1.19    $\pm$  0.10  &   5.67    $\pm$    0.75 &       $\cdots$       &          $\cdots$      &~(1)   \\
145    &  HD 167264    &  10.46 & --1.74 &  B0.5Ia        &   1514   & 0.30 &   0.98   $\pm$   0.13   &  0.466   $\pm$  0.033 &   1.55    $\pm$    0.16 & 0.48   $\pm$   0.09  &          $\cdots$      &~(1)   \\
146    &  HD 167756    & 351.47 &--12.30 &  B0.5Iab?      &   4230   & 0.07 &         $\cdots$        &  0.460   $\pm$  0.005 &   6.57    $\pm$    1.01 &       $\cdots$       &   20.42   $\pm$  12.55 &~(1)   \\
147    &  HD 168076    &  16.94 &  +0.84 &  O5V           &   1820   & 0.76 &   2.64   $\pm$   0.13   &  2.58    $\pm$  0.20  &   3.39    $\pm$    0.31 & 0.98   $\pm$   0.12  &          $\cdots$      &~(1)   \\
148m   &  HD 170740    &  21.06 & --0.53 &  B2V           &    235   & 0.47 &   1.37   $\pm$   0.09   &  2.09    $\pm$  0.20  &   4.45    $\pm$    0.52 & 1.53   $\pm$   0.25  &          $\cdots$      &~(5)   \\
149    &  HD 175360    &  12.53 &--11.29 &  B6III         &    270   & 0.12 &         $\cdots$        &  0.396   $\pm$  0.0329&   3.30    $\pm$    0.55 &       $\cdots$       &          $\cdots$      &~(3)   \\
150    &  HD 177989    &  17.81 &--11.88 &  B2II          &   5021   & 0.22 &   0.63   $\pm$   0.09   &  0.88    $\pm$  0.20  &   4.00    $\pm$    1.09 & 1.40   $\pm$   0.52  &    6.44   $\pm$   4.85 &~(1)   \\
151mv  &  HD 179406    &  28.23 & --8.31 &  B3IVvar       &    227   & 0.31 &   0.89   $\pm$   0.09   &  1.30    $\pm$  0.03  &   4.19    $\pm$    0.23 & 1.46   $\pm$   0.18  &          $\cdots$      &~(7)   \\
152ms  &  HD 184915    &  31.77 &--13.29 &  B0.5IIIne     &    700   & 0.22 &   0.68   $\pm$   0.07   &  1.39    $\pm$  0.331 &   6.32    $\pm$    1.79 & 2.06   $\pm$   0.70  &          $\cdots$      &~(5)   \\
153    &  HD 185418    &  53.60 & --2.17 &  B0.5 V        &   1027   & 0.47 &   1.17   $\pm$   0.09   &  0.74    $\pm$  0.20  &   1.57    $\pm$    0.46 & 0.63   $\pm$   0.22  &    8.86   $\pm$   4.26 &~(1)   \\
154    &  HD 186994    &  78.62 & +10.06 &  B0III         &   2500   & 0.16 &   0.53   $\pm$   0.08   &  0.74    $\pm$  0.20  &   4.63    $\pm$    1.54 & 1.41   $\pm$   0.60  &          $\cdots$      &~(1)   \\
155    &  HD 188209    &  80.99 & +10.09 &  O9.5Ib        &   2210   & 0.15 &   0.68   $\pm$   0.11   &  0.274   $\pm$  0.0288&   1.83    $\pm$    0.31 & 0.40   $\pm$   0.11  &          $\cdots$      &~(3)   \\
156s   &  HD 190918    &  72.65 &  +2.06 &  WN4+O9.7Iab   &   2290   & 0.41 &         $\cdots$        &  0.470   $\pm$  0.032 &   1.15    $\pm$    0.11 &       $\cdots$       &          $\cdots$      &~(1)   \\
157s   &  HD 192035    &  83.33 &  +7.76 &  B0III-IVn     &   2800   & 0.35 &         $\cdots$        &  1.473   $\pm$  0.0559&   4.21    $\pm$    0.28 &       $\cdots$       &          $\cdots$      &~(3)   \\
158iv  &  HD 192639    &  74.90 &  +1.48 &  O8V           &    999   & 0.61 &   1.92   $\pm$   0.13   &  0.2     $\pm$  0.1   &   0.33    $\pm$    0.17 & 0.10   $\pm$   0.06  &    6.97   $\pm$   4.79 &(18)   \\
159    &  HD 195965    &  85.71 &  +5.00 &  B0V           &   1300   & 0.22 &   0.68   $\pm$   0.10   &  1.01    $\pm$  0.20  &   4.59    $\pm$    1.12 & 1.49   $\pm$   0.51  &    9.70   $\pm$  10.08 &~(1)   \\
160    &  HD 197512    &  87.89 &  +4.63 &  B1V           &   1614   & 0.29 &   0.70   $\pm$   0.07   &  1.145   $\pm$  0.0654&   3.95    $\pm$    0.36 & 1.62   $\pm$   0.25  &          $\cdots$      &~(3)   \\
161m   &  HD 198478    &  85.75 &  +1.49 &  B3Ia          &    890   & 0.57 &   1.48   $\pm$   0.13   &  2.68    $\pm$  0.02  &   4.70    $\pm$    0.12 & 1.81   $\pm$   0.18  &          $\cdots$      &~(7)   \\
162    &  HD 198781    &  99.94 & +12.61 &  B2IV          &    768   & 0.31 &   0.80   $\pm$   0.07   &  1.38    $\pm$  0.20  &   4.45    $\pm$    0.79 & 1.73   $\pm$   0.41  &          $\cdots$      &~(1)   \\
163    &  HD 199579    &  87.50 & --0.30 &  B0.5V         &    990   & 0.33 &   1.01   $\pm$   0.09   &  0.88    $\pm$  0.20  &   2.67    $\pm$    0.69 & 0.87   $\pm$   0.27  &          $\cdots$      &~(1)   \\
164    &  HD 201345    &  78.44 & --9.54 &  O9V           &   2570   & 0.17 &         $\cdots$        &  0.183   $\pm$  0.0688&   1.08    $\pm$    0.47 &       $\cdots$       &          $\cdots$      &~(3)   \\
165    &  HD 202347    &  88.22 & --2.08 &  B1V           &   1300   & 0.19 &   0.53   $\pm$   0.08   &  0.271   $\pm$  0.0659&   1.43    $\pm$    0.42 & 0.51   $\pm$   0.21  &   17.75   $\pm$   8.57 &~(3)   \\
166s   &  HD 202904    &  80.98 &--10.05 &  B2Vne         &    276   & 0.13 &         $\cdots$        &  0.33    $\pm$  0.05  &   2.54    $\pm$    0.58 &       $\cdots$       &    7.47   $\pm$  11.56 &~(1)   \\
167s   &  HD 203374    & 100.51 &  +8.62 &  B0IVpe        &    820   & 0.60 &   1.88   $\pm$   0.21   &  0.37    $\pm$  0.20  &   0.62    $\pm$    0.34 & 0.20   $\pm$   0.13  &    7.26   $\pm$   8.34 &~(1)   \\
168m   &  HD 203532    & 309.46 &--31.74 &  B5V           &    211   & 0.28 &   0.94   $\pm$   0.10   &  1.37    $\pm$  0.035 &   4.89    $\pm$    0.30 & 1.45   $\pm$   0.19  &          $\cdots$      &~(5)   \\
169v   &  HD 206267    &  99.29 &  +3.74 &  O6V           &    814   & 0.52 &   1.47   $\pm$   0.11   &  1.27    $\pm$  0.06  &   2.44    $\pm$    0.16 & 0.87   $\pm$   0.11  &    9.29   $\pm$   4.24 &(19)   \\
170    &  HD 206773    &  99.80 &  +3.62 &  B0V           &    597   & 0.45 &   1.99   $\pm$   0.16   &  2.03    $\pm$  0.20  &   4.51    $\pm$    0.54 & 1.02   $\pm$   0.18  &          $\cdots$      &~(1)   \\
171m   &  HD 207198    & 103.14 &  +6.99 &  O9II          &   1216   & 0.54 &   1.50   $\pm$   0.22   &  0.98    $\pm$  0.01  &   1.81    $\pm$    0.05 & 0.66   $\pm$   0.10  &   10.01   $\pm$   3.57 &~(7)   \\
172s   &  HD 207308    & 103.11 &  +6.82 &  B0.7III-IVn   &   1470   & 0.52 &   1.61   $\pm$   0.19   &  0.88    $\pm$  0.20  &   1.69    $\pm$    0.42 & 0.55   $\pm$   0.19  &    9.25   $\pm$   4.15 &~(1)   \\
173mv  &  HD 207538    & 101.60 &  +4.67 &  O9.5V         &    880   & 0.64 &   1.44   $\pm$   0.21   &  2.24    $\pm$  0.03  &   3.50    $\pm$    0.10 & 1.56   $\pm$   0.25  &    9.43   $\pm$   4.16 &~(8)   \\
174    &  HD 208440    & 104.03 &  +6.44 &  B1V           &    620   & 0.34 &         $\cdots$        &  1.287   $\pm$  0.0559&   3.79    $\pm$    0.28 &       $\cdots$       &          $\cdots$      &~(3)   \\
175    &  HD 208947    & 106.55 &  +9.00 &  B2V           &    500   & 0.19 &   0.63   $\pm$   0.18   &  0.51    $\pm$  0.20  &   2.68    $\pm$    1.19 & 0.81   $\pm$   0.55  &          $\cdots$      &~(1)   \\
176    &  HD 209339    & 104.58 &  +5.87 &  B0IV          &    980   & 0.35 &   1.09   $\pm$   0.14   &  1.66    $\pm$  0.20  &   4.74    $\pm$    0.71 & 1.53   $\pm$   0.38  &    6.09   $\pm$   4.48 &~(1)   \\
177m   &  HD 210121    &  56.88 &--44.46 &  B9V           &    223   & 0.38 &   0.76   $\pm$   0.08   &  1.32    $\pm$  0.04  &   3.47    $\pm$    0.20 & 1.73   $\pm$   0.23  &          $\cdots$      &(20)   \\
178iv  &  HD 210809    &  99.85 & --3.13 &  O9Ib          &   3961   & 0.31 &   1.05   $\pm$   0.13   &  0.6     $\pm$  0.1   &   1.94    $\pm$    0.39 & 0.57   $\pm$   0.17  &          $\cdots$      &(18)   \\
179iv  &  HD 210839    & 103.83 &  +2.61 &  O6Iab         &   1260   & 0.57 &   1.15   $\pm$   0.17   &  1.4     $\pm$  0.1   &   2.46    $\pm$    0.22 & 1.22   $\pm$   0.26  &    8.35   $\pm$   4.21 &(18)   \\
180    &  HD 212791    & 101.64 & --4.30 &  B8            &    370   & 0.06 &         $\cdots$        &  1.183   $\pm$  0.0572&  19.72    $\pm$    4.24 &       $\cdots$       &          $\cdots$      &~(3)   \\
\noalign{\smallskip}\hline\noalign{\smallskip}
\end{tabular}
\ec
\end{table}
\end{landscape}
\setcounter{table}{0}
\begin{landscape}
\begin{table}
\caption[]{continued.}
\bc
\begin{tabular}{llrrcrcccrccc}
\hline\hline\noalign{\smallskip}
\multicolumn{1}{l}{~~$N$} &
\multicolumn{1}{l}{~~~Star} &
\multicolumn{1}{c}{$l$} &
\multicolumn{1}{c}{~$b$} &
\multicolumn{1}{c}{Spectrum} &
\multicolumn{1}{c}{$D$, pc} &
\multicolumn{1}{c}{$E(B-V)$} &
\multicolumn{1}{c}{$A_V$} &
\multicolumn{1}{c}{$P,\,\%$} &
\multicolumn{1}{c}{$P/E(B-V),\%$} &
\multicolumn{1}{c}{$P/A_V,\%$} &
\multicolumn{1}{c}{$\left [{\rm {Fe(rest)}/{H}} \right ]_{\rm d}$} &
\multicolumn{1}{c}{Reference} \\
\multicolumn{1}{l}{~(1)} &
\multicolumn{1}{l}{~~~(2)} &
\multicolumn{1}{c}{(3)} &
\multicolumn{1}{c}{~(4)} &
\multicolumn{1}{c}{(5)} &
\multicolumn{1}{c}{(6)} &
\multicolumn{1}{c}{(7)} &
\multicolumn{1}{c}{(8)} &
\multicolumn{1}{c}{(9)} &
\multicolumn{1}{c}{(10)} &
\multicolumn{1}{c}{(11)} &
\multicolumn{1}{c}{(12)} &
\multicolumn{1}{c}{(13)} \\
\noalign{\smallskip}\hline\noalign{\smallskip}
181m   &  HD 214680    &  96.65 &--16.98 &  O9V           &    610   & 0.08 &   0.21   $\pm$   0.08   &  0.55    $\pm$  0.124 &   6.88    $\pm$    2.41 & 2.62   $\pm$   1.58  &   13.96   $\pm$  10.46 &~(9)  \\
182    &  HD 214993    &  97.65 &--16.18 &  B1.5IIIn      &    610   & 0.10 &   0.50   $\pm$   0.13   &  0.467   $\pm$  0.0309&   4.67    $\pm$    0.78 & 0.93   $\pm$   0.30  &   17.12   $\pm$  10.12 &~(3)   \\
183    &  HD 215733    &  85.16 &--36.35 &  B1II          &   2900   & 0.10 &         $\cdots$        &  0.24    $\pm$  0.03  &   2.40    $\pm$    0.54 &       $\cdots$       &   16.44   $\pm$  18.39 &~(1)   \\
184    &  HD 218376    & 109.96 & --0.79 &  B1III         &    383   & 0.23 &   0.71   $\pm$   0.10   &  0.84    $\pm$  0.20  &   3.65    $\pm$    1.03 & 1.18   $\pm$   0.45  &    5.02   $\pm$   6.93 &~(1)   \\
185s   &  HD 218915    & 108.06 & --6.89 &  O9.5Iabe      &   3660   & 0.26 &   0.70   $\pm$   0.39   &  1.08    $\pm$  0.20  &   4.15    $\pm$    0.93 & 1.53   $\pm$   1.13  &          $\cdots$      &~(1)   \\
186    &  HD 219188    &  83.03 &--50.17 &  B0.5III       &   1064   & 0.08 &         $\cdots$        &  0.147   $\pm$  0.0527&   1.84    $\pm$    0.89 &       $\cdots$       &          $\cdots$      &~(3)   \\
187    &  HD 220057    & 112.13 &  +0.21 &  B2IV          &   1421   & 0.24 &   0.56   $\pm$   0.08   &  0.75    $\pm$  0.20  &   3.13    $\pm$    0.96 & 1.34   $\pm$   0.54  &          $\cdots$      &~(1)   \\
188    &  HD 224151    & 115.44 & --4.64 &  B0.5II-III    &   1355   & 0.42 &   1.45   $\pm$   0.16   &  0.567   $\pm$  0.120 &   1.35    $\pm$    0.32 & 0.39   $\pm$   0.13  &    4.73   $\pm$   5.03 &~(1)   \\
189    &  HD 224572    & 115.55 & --6.36 &  B1V           &    340   & 0.19 &   0.51   $\pm$   0.17   &  1.255   $\pm$  0.119 &   6.61    $\pm$    0.97 & 2.45   $\pm$   1.05  &   19.06   $\pm$   8.75 &~(1)   \\
190    &  HD 232522    & 130.70 & --6.71 &  B1II          &   5438   & 0.21 &   0.64   $\pm$   0.12   &  1.61    $\pm$  0.18  &   7.67    $\pm$    1.22 & 2.51   $\pm$   0.74  &          $\cdots$      &~(1)   \\
191v   &  HD 303308    & 287.59 & --0.61 &  O3V           &   3631   & 0.45 &   1.36   $\pm$   0.12   &  3.10    $\pm$  0.12  &   6.89    $\pm$    0.42 & 2.28   $\pm$   0.30  &    4.24   $\pm$   5.53 &(13)   \\
192m   &  HD 308813    & 294.79 & --1.61 &  O9.5V         &   2398   & 0.28 &         $\cdots$        &  1.85    $\pm$  0.03  &   6.61    $\pm$    0.34 &       $\cdots$       &          $\cdots$      &(21)   \\
193s   &  BD +35 4258  &  77.19 & --4.74 &  B0.5 Vn       &   3093   & 0.25 &   0.72   $\pm$   0.08   &  0.51    $\pm$  0.18  &   2.04    $\pm$    0.80 & 0.70   $\pm$   0.33  &    4.34   $\pm$   6.57 &~(1)   \\
194s   &  BD +53 2820  & 101.24 & --1.69 &  B0IV:n        &   4506   & 0.37 &   1.08   $\pm$   0.09   &  2.30    $\pm$  0.18  &   6.22    $\pm$    0.65 & 2.13   $\pm$   0.34  &          $\cdots$      &~(1)   \\
195v   &  CPD -59 2603 & 287.59 & --0.69 &  O7V           &   2630   & 0.46 &   1.45   $\pm$   0.17   &  1.97    $\pm$  0.13  &   4.28    $\pm$    0.38 & 1.36   $\pm$   0.25  &   25.27   $\pm$   9.16 &(13)   \\
196s   &  CPD -69 1743 & 303.71 & --7.35 &  B1Vn          &   4700   & 0.30 &         $\cdots$        &  0.88    $\pm$  0.10  &   2.93    $\pm$    0.43 &       $\cdots$       &          $\cdots$      &~(1)   \\
\noalign{\smallskip}\hline\noalign{\smallskip}
\multicolumn{13}{@{}l@{}}{{\rm Notes.}
`v' --- variable star;
`s' --- star with peculiar spectrum;
`m' --- maximum polarization $P_{\max}$
(wavelength dependence of polarization is
known); `i' --- interstellar polarization;}\\
\multicolumn{13}{@{}l@{}}{References for observational data:
  (1) Heiles (\cite{hei00});
  (2) Ghosh et al. (\cite{ghosh99});
  (3) this paper;
  (4) Anderson \& Wannier (\cite{aw97});
  (5) Serkowski et al. (\cite{smf75});
  (6) Roche et al. (\cite{{roche97}});
  }\\
\multicolumn{13}{@{}l@{}}{
  (7) Efimov (\cite{ef09});
  (8) Martin et al. (\cite{mar99});
  (9) Mcdavid (\cite{mc00});
  (10) Leroy \& Le Borgne (\cite{ll87});
  (11) Oudmaijer  \& Drew (\cite{od99});
  (12) Vink et al. (\cite{vink09});
  (13) Marracco et al. (\cite{mar93});
  }\\
\multicolumn{13}{@{}l@{}}{
  (14) Berdyugin et al. 2010, in preparation;
  (15) Weitenbeck (\cite{wei08});
  (16) Das et al. (\cite{dvi10});
  (17) Poeckert et al.  (\cite{poe79});
  (18) Harries et al. (\cite{har02});
  (19) Elias et al. (\cite{elias08});
  }          \\
\multicolumn{13}{@{}l@{}}{
  (20) Larson et al. (\cite{lars96});
  (21) Vega et al. (\cite{vega94}).} \\
\end{tabular}
\ec
\end{table}
\end{landscape}
\end{document}